\documentclass[aps,prd,nofootinbib,twocolumn,superscriptaddress,preprintnumbers,balancelastpage,longbibliography]{revtex4-1}

\usepackage{placeins}
\usepackage{amsmath,amssymb,mathtools,bm}
\usepackage{graphicx, color, hepunits}
\usepackage[dvipsnames]{xcolor}
\usepackage{cancel}
\usepackage[normalem]{ulem}
\usepackage{float}
\usepackage{filecontents}
\usepackage{multirow}
 \usepackage{hyperref} %Automatically links \label and \ref commands; Always load last
\hypersetup{
    colorlinks=true,       % false: boxed links; true: colored links
    linkcolor=blue,        % color of internal links
    citecolor=blue,        % color of links to bibliography
    filecolor=magenta,     % color of file links
    urlcolor=blue          % color of external links
}
\usepackage[utf8]{inputenc}
\usepackage[english]{babel}

\newcommand{\es}[2] {\begin{equation} \label{#1} \begin{split} #2 \end{split} \end{equation}}

\begin{document}

\title{Detecting axion dark matter beyond the magnetoquasistatic approximation}

\author{Joshua N. Benabou}
\affiliation{Berkeley Center for Theoretical Physics, University of California, Berkeley, CA 94720, U.S.A.}
\affiliation{Theoretical Physics Group, Lawrence Berkeley National Laboratory, Berkeley, CA 94720, U.S.A.}

\author{Joshua W. Foster}
\affiliation{Center for Theoretical Physics, Massachusetts Institute of Technology, Cambridge, Massachusetts 02139, U.S.A}

\author{Yonatan~Kahn}
\affiliation{Illinois Center for Advanced Studies of the Universe and Department of Physics, University of Illinois Urbana-Champaign, Urbana, Illinois 61801, U.S.A.}

\author{Benjamin R. Safdi}
\affiliation{Berkeley Center for Theoretical Physics, University of California, Berkeley, CA 94720, U.S.A.}
\affiliation{Theoretical Physics Group, Lawrence Berkeley National Laboratory, Berkeley, CA 94720, U.S.A.}

\author{Chiara~P.~Salemi}
\affiliation{Laboratory of Nuclear Science, Massachusetts Institute of Technology, Cambridge, MA 02139, U.S.A.}
\affiliation{Kavli Institute for Particle Astrophysics and Cosmology, Stanford University, Stanford, CA 94305, U.S.A.}
\affiliation{Stanford Linear Accelerator Center, Menlo Park, CA 94025, U.S.A.}

\date{\today}
\preprint{MIT-CTP/5490}

\begin{abstract}
A number of proposals have been put forward for detecting axion dark matter (DM) with grand unification scale decay constants that rely on the conversion of coherent DM axions to oscillating magnetic fields in the presence of static, laboratory magnetic fields.  Crucially, such experiments -- including ABRACADABRA -- have to-date worked in the limit that the axion Compton wavelength is larger than the size of the experiment, which allows one to take a  magnetoquasistatic (MQS) approach to  modeling the axion signal.
We use finite element methods to solve the coupled axion-electromagnetism equations of motion without assuming the MQS approximation.
We show that the MQS approximation becomes a poor approximation at frequencies two orders of magnitude lower than the naive MQS limit. Radiation losses diminish the quality factor of an otherwise high-$Q$
resonant readout circuit, though this may be mitigated through shielding and minimizing lossy materials.
Additionally, self-resonances associated with the detector geometry change the reactive properties of the pickup system, leading to two generic features beyond MQS: there are frequencies that require an inductive rather than capacitive tuning to maintain resonance, and the detector itself becomes a multi-pole resonator at high frequencies. Accounting for these features, competitive sensitivity to the axion-photon coupling
may be extended well 
beyond the naive MQS limit.
\end{abstract}
\maketitle

The quantum chromodynamics (QCD) axion with decay constant near the Grand Unification Theory (GUT) scale provides a compelling dark matter (DM) candidate~\cite{Preskill:1982cy,Abbott:1982af,Dine:1982ah}, a solution to the strong-CP problem from the non-observation of a neutron electric dipole moment~\cite{Peccei:1977hh,Peccei:1977ur,Weinberg:1977ma,Wilczek:1977pj}, and may emerge naturally in ultraviolet theories from String Theory~\cite{Witten:1984dg,Svrcek:2006yi,Arvanitaki:2009fg,Halverson:2019cmy} and GUT field theories~\cite{Wise:1981ry,Ballesteros:2016xej,Ernst:2018bib,DiLuzio:2018gqe,Ernst:2018rod,FileviezPerez:2019fku,FileviezPerez:2019ssf,Co:2016xti}.   The axion is naturally realized as the Goldstone boson of a $U(1)$ symmetry, called the Peccei-Quinn (PQ) symmetry, that is spontaneously broken at a high energy scale $f_a$~\cite{DiLuzio:2020wdo}.  The axion acquires a non-trivial potential from QCD instantons, allowing it to solve the strong-CP problem, and also leading to an axion mass $m_a \approx 0.57 \, \, {\rm neV} (10^{16} \, \, {\rm GeV} / f_a)$~\cite{diCortona:2015ldu}.  

A number of compelling experiments have been proposed for detecting GUT-scale axion DM in the laboratory that rely on the coupling of the axion $a$ to electromagnetism ${\mathcal L} \supset g_{a\gamma\gamma} a {\bf E} \cdot {\bf B}$, with ${\bf E}$ (${\bf B}$) the electric (magnetic) field and $g_{a\gamma\gamma}$ the axion-photon coupling (see~\cite{Adams:2022pbo} for a review).  Axion DM behaves as a classical wave, whose time dependence is $a(t) \approx a_0 \cos(m_a t)$
with amplitude $a_0$ set by the local DM density $\rho_{\rm DM}$: $ m_a^2 a_0^2 / 2 = \rho_{\rm DM}$~\cite{Foster:2017hbq}.   The axion field may convert to electromagnetic waves with frequency $\omega \approx m_a$ in the presence of static, external magnetic fields; for axion masses $m_a \sim 5$ $\mu$eV (corresponding to $f_a \sim 10^{12}$ GeV and Compton wavelength $\sim$25 cm), that radiation
may then be enhanced in a resonant cavity of comparable size~\cite{Sikivie:1983ip}. The ADMX~\cite{ADMX:2019uok,ADMX:2018gho,ADMX:2021nhd} and HAYSTAC~\cite{Brubaker:2016ktl,HAYSTAC:2018rwy,HAYSTAC:2020kwv} experiments, amongst others~\cite{Adams:2022pbo}, have successfully searched for QCD axion DM in this mass range using resonant cavities.  The problem with using resonant cavity experiments to search for GUT-scale axion DM is clear: probing $m_a \sim 0.5$ neV would require a 
cavity with size $\sim$2.5 km.

However, in the limit where the axion Compton wavelength is much larger than the size of the experiment, the equations of axion-electrodynamics can be solved in the magnetoquasistatic (MQS) approximation, in which case a static laboratory magnetic field ${\bf B}_0$ sources an effective current ${\bf J}_{\rm eff}(t) \approx g_{a\gamma\gamma} \sqrt{2 { \rho_{\rm DM}}} \cos(m_a t) {\bf B}_0$~\cite{Sikivie:2013laa,Kahn:2016aff}.  The effective current oscillates in time, creating a real, secondary, oscillating magnetic field which can be used to drive current in a pickup loop via Faraday's law. Such detectors, which include ABRACADABRA-10 cm (ABRA-10 cm)~\cite{Ouellet:2018beu,PhysRevD.99.052012,Salemi:2021gck}, SHAFT~\cite{Gramolin:2020ict}, and ADMX SLIC~\cite{Crisosto:2019fcj},  are generically referred to as ``lumped element" detectors, since in the MQS approximation they may be described through lumped-element circuit components such as inductors and capacitors~\cite{Adams:2022pbo}.  

In this Letter we study how detectors designed to operate in the MQS regime behave when the axion Compton wavelength approaches the size of the detector. It is clear that the MQS approximation is applicable for $m_a^{-1} \gg L$, with $L$ the characteristic size of the detector, but it is unclear precisely where the MQS approximation breaks down and how this affects the sensitivity of planned lumped-element detectors. We note that most detector designs to-date have have been informed by the  MQS calculation at masses up to $m_a^{-1} \sim L$~\cite{Kahn:2016aff,Adams:2022pbo,Brouwer:2022bwo,DMRadio:2022pkf};
in this work, however, we show that the MQS assumption receives important corrections at much lower frequencies, which provides insights that may influence the design of future detectors.
For example, the upcoming DMRadio program~\cite{Adams:2022pbo,Brouwer:2022bwo,DMRadio:2022pkf}, will feature both meter-sized (DMRadio-50 L and DMRadio-m$^3$) and larger (DMRadio-GUT) experiments using solenoidal and toroidal magnets with LC resonant readouts~\cite{Chaudhuri:2018rqn,Chaudhuri:2019ntz,MethodsPaper} to probe QCD axion DM 
across the entire mass range from $\sim$0.4 neV to nearly 1 $\mu$eV.

\begin{figure}[!t]
    \centering
    \includegraphics[width=1\columnwidth]{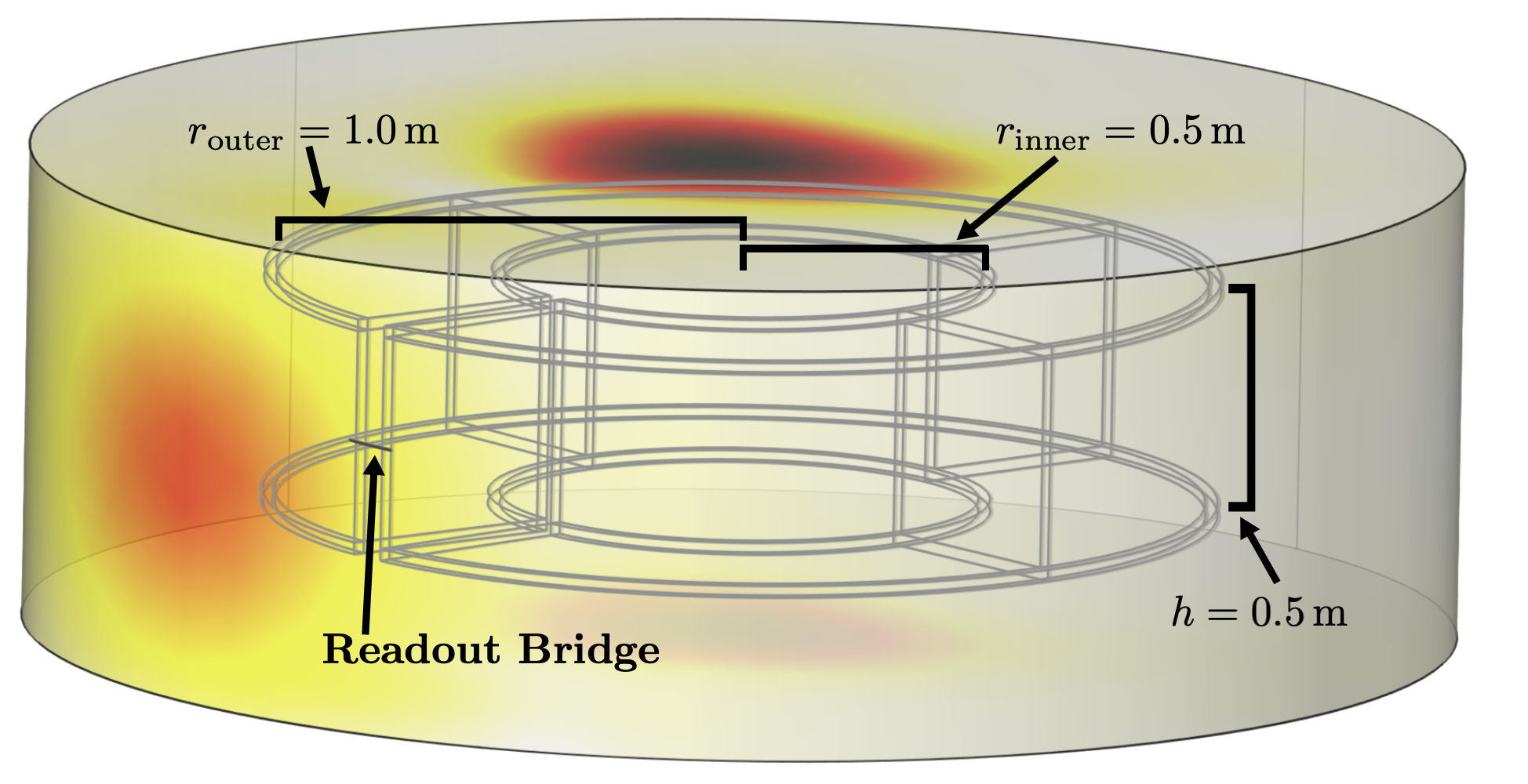}
    \caption{An illustration of our fiducial toroid with inner radius $0.5$ m, outer radius $1$ m, and height $0.5$ m, along with a 5$^\circ$ angular gap.  A 10 T toroidal magnetic field sources an axion effective current, which generated a magnetic flux that pierces the center of the toroid, inducing a current that flows along the superconducting sheath that surrounds the toroid.  A lumped element port is added to the wire bridging the gap, where the adjustable capacitor (for achieving resonance) is located along with the inductive coupling to the SQUID amplifier circuit. The heatmap shows the Poynting vector flux through 
    a surface far from the detector 
    for a simulation in the unshielded configuration for a $f=50$ MHz signal.
    }
    \label{fig:Toroid}
\end{figure}

\noindent
{\bf Analytic considerations past the MQS limit.---}  
We consider a toy detector roughly modeled on the ABRA proposals in~\cite{Kahn:2016aff} with resonant readout, as illustrated in Fig.~\ref{fig:Toroid} (see the Supplementary Material (SM) for solenoidal geometry and broadband readout results).  The inner toroid radius is $r_{\rm inner}$, the outer radius is $r_{\rm outer}$, the height is $h$, and the angular gap size where the readout is located is $\theta_{\rm gap}$.  
We assume that a constant magnetic field of magnitude $B_0$ fills the toroid and gap. 
The toroid is surrounded by a superconducting sheath with the exception of the gap; the gap is connected by a wire as illustrated in Fig.~\ref{fig:Toroid}. In the middle of the wire, there is an inductive coupling to the SQUID amplifier readout and also additional lumped-element circuit components tuned to achieve resonance at the frequency of interest.  

In the MQS limit ($m_a \ll r_{\rm outer}^{-1}, \, h^{-1}$) the sensitivity of the detector to $g_{a\gamma\gamma}$ may be estimated as follows~\cite{Kahn:2016aff}.  Let us assume that there is an axion signal with mass $m_a$ and that a capacitor has been added to the readout such that there is a resonance in the LC circuit, with the toroid providing the inductance, at $\omega = m_a$, with quality factor $Q \lesssim 10^6$. Since the axion signal has a bandwidth $\delta f / f \lesssim 10^{-6}$, we may assume for simplicity that the entire signal is within the full-width half max (FWHM) of the LC circuit. (Quality factors larger than $10^6$ would be optimal but introduce additional complications that are not important for this discussion \cite{Chaudhuri:2018rqn}.)  Furthermore, we assume that thermal noise at temperature $T_{\rm LC}$ in the LC circuit and amplifier noise in the DC-SQUID readout, which is optimally inductively coupled to the pickup circuit, are the limiting sources of noise.
Assuming a small gap size, the oscillating axion-induced effective current induces a voltage across the sheath gap, which in the MQS limit we can model as a series RLC circuit. We refer to the power spectral density (PSD) of the induced voltage as the source voltage PSD $S_{VV}^a(\omega)$, the mean of which is given  by~\cite{Foster:2017hbq}
\es{eq:SVV}{
S_{VV}^a(\omega) = \pi \omega^2 g_{a\gamma\gamma}^2 B_0^2 V_{\rm eff}^2 \rho_{\rm DM} f_{\rm DM}(\omega) \,,
}
where $f_{\rm DM}(\omega)$ is the DM velocity distribution ({\it e.g.}, the Standard Halo Model~\cite{Foster:2017hbq}), translated to a normalized frequency distribution, and $V_{\rm eff}$ is an effective volume that depends on the geometry and the pickup system inductance $L_p$, with typical value $\mathcal{O}(0.1)$ times the physical toroid volume.
Note that $f_{\rm DM}(\omega)$ vanishes for $\omega < m_a$ and only has nontrivial support up to $\omega \sim m_a (1 + 10^{-6})$, given the Galactic DM velocities.
On resonance $(\omega = m_a)$, Eq.~(\ref{eq:SVV}) induces a current PSD $S_{II}^a = S_{VV}^a \big[ Q  / (m_a L_p)\big]^2$, with $Q$ the quality factor of the circuit. 
In contrast, the current PSD from thermal noise (at high occupation numbers) and amplifier noise at resonance with an optimally coupled amplifier (see~\cite{Chaudhuri:2018rqn})
is $S_{II}^{\rm noise} \approx Q (2 T_{\rm LC} + \eta_A \omega) / (\pi m_a L_p)$. Here $\eta_A$ is a factor that parameterizes how far away the amplifier is from the standard quantum limit (SQL), with $\eta_A = 1$ being at the SQL; in the main Letter we assume $\eta_A = 20$, as targeted by DM Radio-m$^3$~\cite{DMRadio:2022pkf}, though in the SM we show figures with $\eta_A = 1$.  The  precise value of $\eta_A$ does not qualitatively affect our discussion.

The signal and noise PSDs may then be incorporated into a likelihood analysis~\cite{Foster:2017hbq}, the result of which is a projected sensitivity characterized by test statistic $\Theta =  t_{\rm exp}/2\pi \times \int d \omega \left( S^a_{II} / S^{\rm noise}_{II} \right)^2$ with $t_{\rm exp}$ the  data taking time, which we assume is large enough such that the signal is resolved by multiple frequency bins. 
If we are interested in scanning over {\it e.g.} one decade of possible masses starting at some lower frequency $m_a^0$ with a total exposure time $t_{\rm tot}$, then, assuming a scanning strategy where we follow the QCD band $g_{a\gamma\gamma} \propto m_a$, the  amount of time $t_{\rm exp}(m_a)$ we spend at a given frequency $m_a > m_a^0$ scales with mass as 
$t_{\rm exp}(m_a) \propto (m_a^0 / m_a)^n$, where $n = 5$ when dominated by thermal noise and $n = 3$ when dominated by readout noise \cite{Chaudhuri:2018rqn, Chaudhuri:2019ntz}.

\begin{figure}[!t]
    \centering
    \includegraphics[width=1\columnwidth]{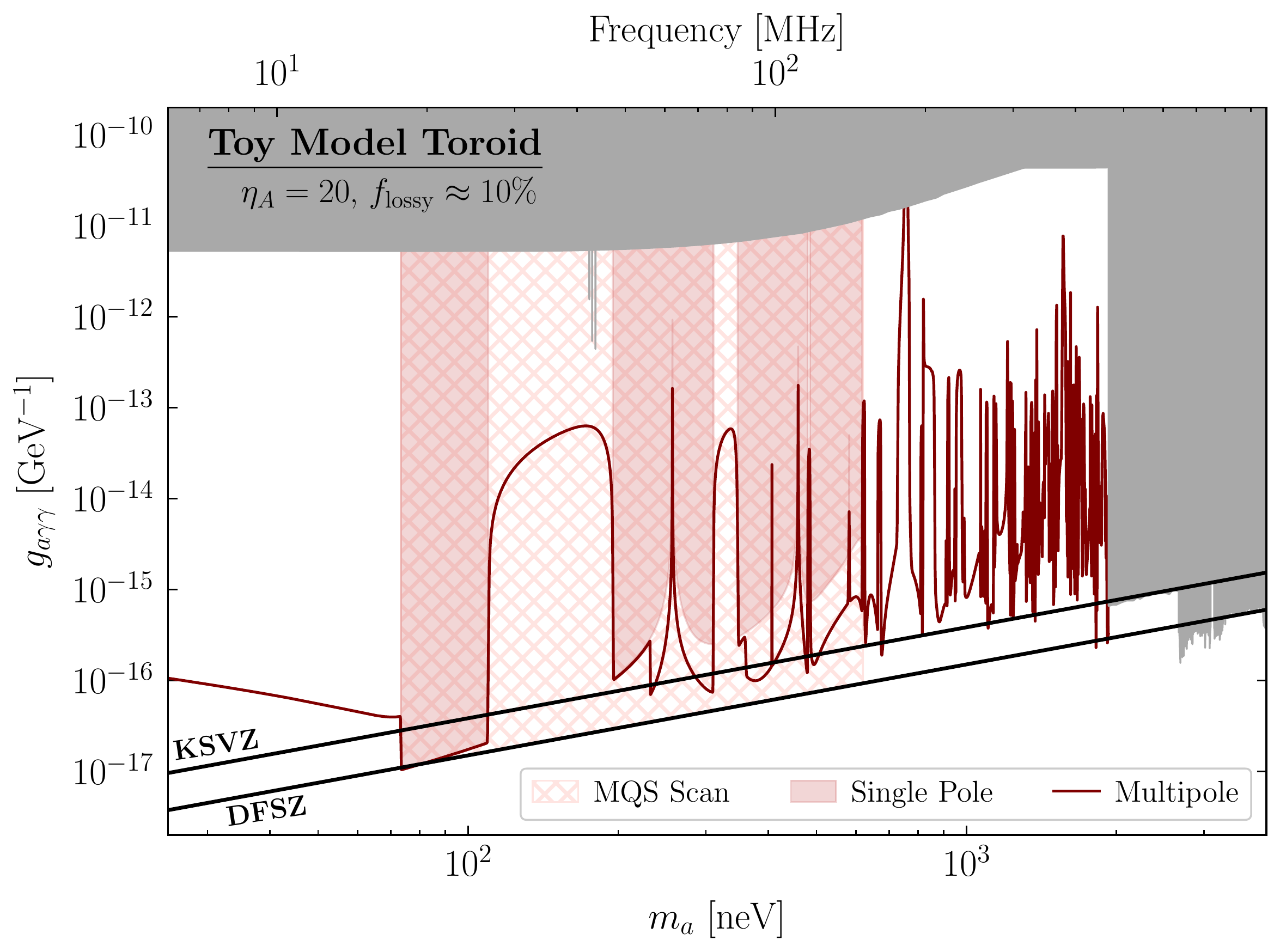}
    \caption{ The projected sensitivity of our fiducial toroid
    to $g_{a\gamma\gamma}$ at 95\% confidence as a function of axion mass $m_a$ for the scenario of a perfectly reflecting, surrounding shield and 10\% by volume lossy material within the toroid.  The MQS expectation is shown hatched, with a scanning strategy that maintains sensitivity to the DFSZ axion starting from the naive MQS breakdown frequency $f_{\rm MQS}$ and going  to lower masses over a one year time scale.  The same scanning strategy without the MQS approximation, as simulated using \texttt{COMSOL}, gives the sensitivity illustrated by ``Single Pole," which makes use of  a single pole readout strategy without inductive tunings. Using the full high frequency response at each tuning, accounting for the multipolar response of the resonant system at high frequencies, yields the sensitivity curve labelled ``Multipole," which extends the mass range of the lumped element detector all the way to masses probed by ADMX.}
    \label{fig:sensitivity}
\end{figure}

As an illustration, we consider throughout this Letter a toroid, which we call our fiducial detector, with dimensions $r_{\rm outer} = 2 r_{\rm inner} = 2 h = 1$ m, with $B_0 = 10$ T.  The pickup sheath has inductance $L_p \approx \pi r_{\rm inner}^2 / h$. 
An adjustable capacitor is added to the lumped-element port in series with the pickup circuit, and the resonant frequency is tuned to search for axion DM starting at an upper frequency that we call the putative MQS breakdown frequency, $f_{\rm MQS} =1/( 4  r_{\rm inner}) \approx 150$ MHz, which is the inverse of the diameter of the toroid.  For definiteness we assume one year of total data taking time, and we implement a search strategy where we scan in axion mass from $m_a = 2 \pi f_{\rm MQS}$ to lower masses, maintaining sensitivity to the DFSZ axion~\cite{Dine:1981rt,Zhitnitsky:1980tq}.  The lowest frequency we reach in this search is $f \approx 17$ MHz, as illustrated in Fig.~\ref{fig:sensitivity}. 
Note that to simplify the discussion we assume that at all resonant frequencies the LC circuit has a quality factor $Q = 10^6$.
We also assume that $T_{\rm LC} = 20$ mK and the readout noise parameter is $\eta_A = 20$.

Going beyond the MQS approximation, we need to solve Maxwell's equations coupled to the axion source ${\bf J}_{\rm eff}$ with the boundary conditions specified by the detector.
Note that there are also contributions from axion field gradients, but such terms are subdominant relative to ${\bf J}_{\rm eff}$ by factors of the DM velocity, so we neglect them in this analysis.  The MQS approximation amounts to neglecting the time-derivative terms in Maxwell's equations, as well as any retarded-time effects.

To begin, we work to leading order in frequency, where we may still treat the pickup loop circuit in the lumped-element approximation and neglect finite propagation time effects.  However, an oscillating (effective) current source will radiate. To keep track of the radiative power losses (otherwise known as radiation resistance), we first assume that there is no surrounding shield (in the ultra-low frequency limit the shield plays no role in determining the quality factor).  We may describe the radiation power from the toroid in the dipole approximation. For illustrative purposes, we assume the toroid has two independent geometric scales, $r_{\rm outer} = 2 r_{\rm inner} = 2 r$ and $h$ independent of $r$. Approximating the surface currents on the toroidal surface by a uniform current density $J$ through the inner toroidal volume, the time-averaged radiated power is $P_{\rm rad} = m_a^4 G^2 J^2 / (12\pi)$, where $G = (7 \pi / 3) h r^3$ is a geometric factor. We may then identify a radiation resistance $R_{\rm rad}$ by setting $P_{\rm rad} = R_{\rm rad} I^2 / 2$, with $I$ the linear current through the cross-section of the toroid.
Then, the total quality factor of the circuit is $Q_{\rm tot} = (1/Q_{\rm prim} + 1/Q_{\rm rad})^{-1}$, with $Q_{\rm prim} \approx 10^6$ the original quality factor before radiative energy loss was accounted for and $Q_{\rm rad} \equiv m_a L_p / R_{\rm rad}$ the quality factor associated with the radiation resistance. A straightforward calculation yields 
\es{eq:Q_an}{
Q_{\rm rad} = \frac{54}{49} \left( \frac{1}{m_a h} \right) \left( \frac{1}{m_a r} \right)^2 \,.
}
Let us now assume that there is a single scale, with $h = r$. Then, to avoid diminishing the total quality factor below $10^6$ we need $m_a \lesssim r^{-1} / 100$. That is, the MQS approximation breaks down two orders of magnitude before $m_a \sim 2\pi f_{\rm MQS}$.
For example, for our fiducial toroid ($f_{\rm MQS} \approx 150$ MHz)
radiation-induced $Q$ degradation becomes important for $f \gtrsim 1$ MHz.

As $m_a r$ increases, the next important effect is that the toroid pickup sheath stops behaving like a lumped-element inductor at its first self-resonant frequency ($f_\mathrm{SRF}$).  Considering the sheath as a transmission line, we expect its reactance $X$ to depend on wavelength $\lambda$ by $X \propto \tan \left( {2 \pi \ell_{\rm perim} / \lambda} \right)$, where $\ell_{\rm perim} \sim 4 \pi r$ is the characteristic perimeter length of the toroid (assuming $r_{\rm outer} = 2 r$). Importantly, this implies that the reactance changes sign when $\lambda \sim 4 \ell_{\rm perim}$, which occurs for $m_a \sim (4 r)^{-1}$.  That is, the pickup loop only behaves like a lumped-element inductor for $m_a \lesssim 0.25 r^{-1}$.  For our fiducial toroid we thus expect $f_{\rm SRF} \sim 25$ MHz, and above this frequency the lumped element approximation is not valid. In particular, the reactance changes sign above $f_{\rm SRF}$, which means that the pickup sheath has capacitive reactance and resonance can only be achieved by adding an additional, tunable lumped-element \emph{inductor} to the readout circuit.
Extending to even higher frequencies, the reactance  oscillates between capacitive and inductive, and the detector acquires multiple poles.

\noindent
{\bf Numerical simulations past the MQS limit.---}Understanding the response of the detector at frequencies beyond $f_{\rm SRF}$ requires numerical simulations of the axion-electrodynamics equations.
Using the RF Module of \texttt{COMSOL} Multiphysics\textsuperscript{\textregistered}~\cite{multiphysics1998introduction}, we simulate our fiducial toroid geometry
including $2.5$ cm thick walls and a $5^\circ$ gap. The toroid is idealized as a perfect electric conductor, and the gap is bridged by a perfectly electrically conducting surface containing a small lumped port, which is then in series with the macroscopic detector element. 

We consider three boundary conditions: boundary conditions at infinity, approximated with perfectly matched layers; perfectly reflecting (\textit{e.g.}, shielded) boundary conditions a finite distance from the detector; and perfectly reflecting boundary conditions with the inclusion of a small amount of absorbing material (plastic) within the detector volume. The unshielded case with boundary conditions at infinity does not correspond to a realistic experimental setup but is useful for illustrating 
radiation resistance.
The reflecting boundary conditions are implemented through a cylindrical, fully-enclosed superconducting cavity with a radius of $1.5$ m and height of 1 m.  In the case where we add absorbing material, we fill the center of the toroid (in the volume containing the magnetic field) with 10\% by volume of TACHYON\textsuperscript{\textregistered} 100G
Ultra Low Loss Laminate and Prepreg, motivated by ABRA-10 cm~\cite{PhysRevD.99.052012} in the sense that some amount of non-superconducting support structure is necessary for the magnet. 
In the SM we show results for larger absorbing fractions. In addition, non-superconducting material may be present to cool the magnet; ABRA-10 cm, for example, wrapped the toroid in copper straps to help with thermalization. Our choice of absorbing material illustrates how the volume of absorbing material is an important design parameter for future detectors; it is not meant as a realistic reflection of the parameters for upcoming experiments such as DMRadio-m$^3$~\cite{MethodsPaper}.    

As in the MQS calculation, we characterize the detector response through an equivalent circuit with the source-voltage PSD $S^a_{VV}(\omega)$ in series with a source impedance $Z_S$ associated with  the detector (see~\cite{MethodsPaper} for details).
To determine the source impedance, the axion effective current is turned off and the lumped port is chosen to be a series frequency-dependent voltage source.
A measurement of the current which flows through the lumped port enables a direct calculation of the equivalent source impedance of the detector. 
To determine the equivalent source voltage, we restore the axion effective current, remove the lumped port source voltage, and measure the voltage drop across the lumped port resistor.  
The equivalent circuit then gives the response of 
the system  when in series with an arbitrary load.
\begin{figure}[!t]
    \centering
    \includegraphics[width=.99\columnwidth]{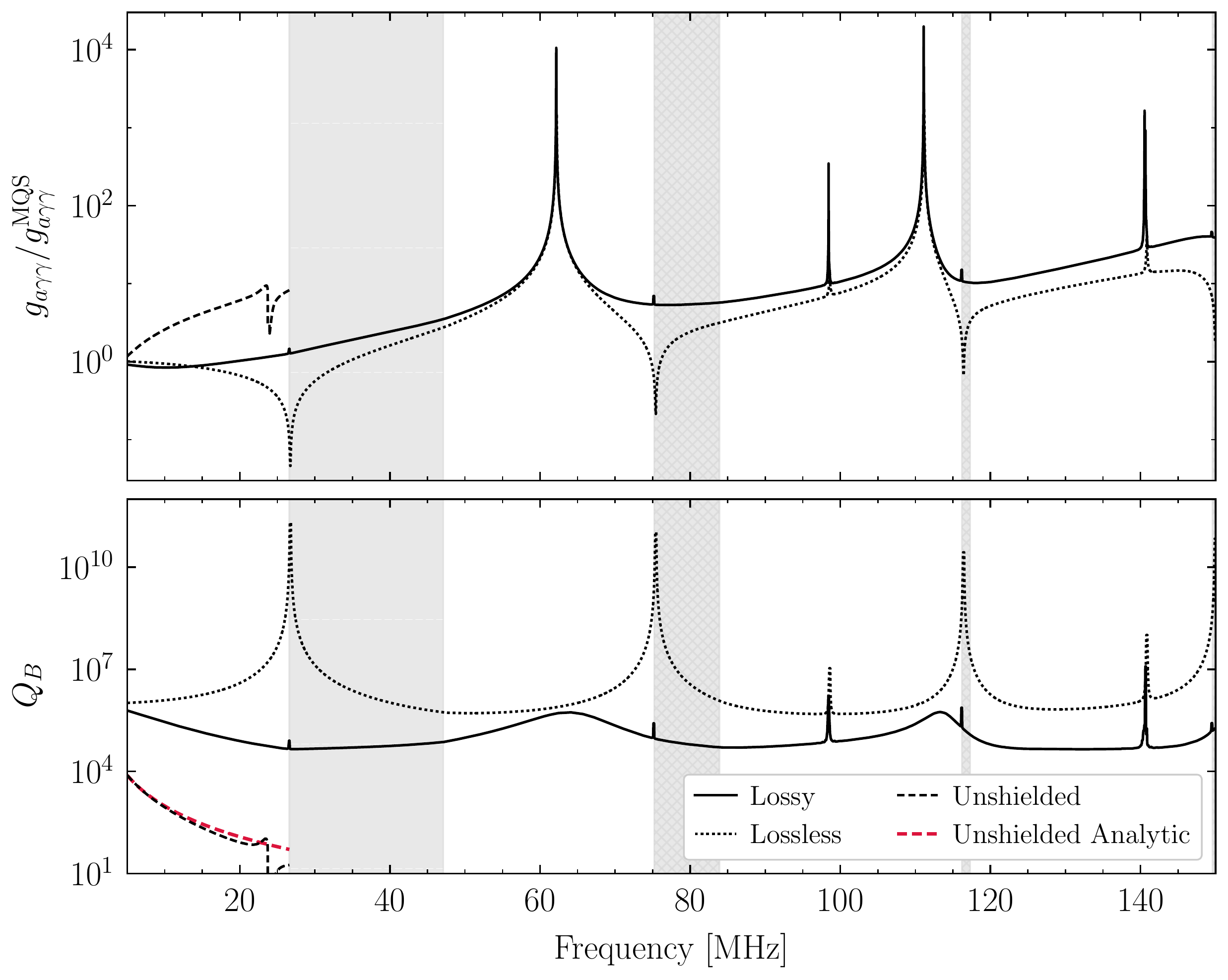}
    \caption{(Top) A comparison of the expected sensitivity of our toy detector geometry for the three boundary conditions considered in this work relative to the MQS expectation (which is independent of assumed boundary conditions). Shaded grey regions indicate frequency ranges that require an inductive tuning for the lossy scenario. (Bottom) The quality factor $Q_B$, which determines the bandwidth of the single-pole response for the different scenarios, along with the low-frequency analytic expectation for $Q_B$.
    }
    \label{fig:rel_sensitivity}
\end{figure}

In the bottom panel of Fig.~\ref{fig:rel_sensitivity}, we show the quality factors measured in the simulation for the three different shielding and absorbing material scenarios.
More precisely, we define the frequency-dependent quantity $Q_B \equiv \omega / \delta \omega$, with  $\delta \omega$ the numerically-measured FWHM of the response about resonance.  Recall that in the MQS approximation we would have $Q_B = 10^6$ by construction across all frequencies. 
With no shield, the quality factor follows the analytic expectation from dipole radiation derived in~\eqref{eq:Q_an}.  The dipole formula is expected to break down at $f_{\rm SRF}$, and this may be clearly seen in Fig.~\ref{fig:rel_sensitivity}.

Note that at frequencies directly above $f_{\rm SRF}$ (shaded) achieving resonance requires the addition of an inductive lumped element component at the readout port, which may be difficult to achieve in practice.  Above $f_{\rm SRF}$ the quality factor is less than $\sim$10 at virtually all frequencies with no shield, due to radiation resistance. The inclusion of a perfectly reflecting shield restores $Q_B$ across all frequencies. We note that with the perfect shield, $Q_B$ formally diverges at $f_{\rm SRF}$.  This is because the resistance, which is added to give finite $Q = 10^6$, is added to the lumped port; at $f_{\rm SRF}$ there is effectively a zero-resistance LC circuit in parallel with the lumped port, meaning that the resistance in the lumped port does not set the quality factor of the resonator. 
Next, we keep the perfectly reflecting shield but add in the lossy material, as described above, to the inside of the toroid.  
The quality factor is slightly degraded as a function of frequency, though it remains $\gtrsim 10^4$ across all frequencies shown. 

We combine the calculation of the quality factor with that of the source voltage that is estimated with our numerical simulations in \texttt{COMSOL} to compute the on-resonance sensitivity to $g_{a\gamma\gamma}$,
and we compare to the sensitivity estimated in the MQS approximation (see~\cite{MethodsPaper} for further details). This comparison is made under the assumption of an intrinsic frequency-dependent resistance such that $Q_B = 10^6$ at all frequencies in the MQS approximation, and with readout noise (parameterized by $\eta_A$) independently tuned to minimize noise on resonance for each of the scenarios \cite{Chaudhuri:2018rqn}.  The sensitivity ratio $g_{a\gamma\gamma} / g_{a\gamma\gamma}^{\rm MQS}$ is illustrated in the top panel of Fig.~\ref{fig:rel_sensitivity} for our different shielding and loss assumptions (we do not show the unshielded case above $f_{\rm SRF}$).  

Even in the ideal scenario, with no loss and perfectly reflecting boundary conditions, at frequencies above $f_{\rm SRF}$ the sensitivity of the detector is degraded relative to the expectation under the MQS approximation due to a decrease in the source voltage relative to the MQS approximation. Moreover, as illustrated in shaded grey, much of the frequency range above $f_{\rm SRF}$ requires inductive tuning.  On the other hand, performing capacitive tuning up to $f_{\rm SRF}$ naturally covers, for free, much of the high-frequency parameter space since the response has multiple poles at high frequency.  Note that in Fig.~\ref{fig:rel_sensitivity} we account for the fact that when the quality factor drops below $10^6$ the scanning may be performed more coarsely, which allows for more integration time per mass point relative to the MQS strategy.  Similarly, when $Q_B \gg 10^6$
only a fraction of the signal is amplified.

In Fig.~\ref{fig:sensitivity}, we show the result of our numerical simulation 
for our fiducial detector in the shielded but lossy material scenario (labeled ``Single Pole").
We implement a one-year scan in which resonant tunings are performed at frequencies with a relative step size $\delta f/f = 1/Q_B^\mathrm{MQS} = 10^{-6}$, and data is collected at each tuning until the sensitivity on resonance would reach the DFSZ benchmark under the MQS approximation.  (Note that as shown in \cite{Chaudhuri:2018rqn, Chaudhuri:2019ntz}, more optimal scanning strategies may extend the range of masses at which DFSZ benchmark sensitivity may be achieved.) Unlike in the MQS approximation (hatched region), some frequencies in the full numerical response would require an inductive load to achieve resonance; these frequencies are excluded from our scanning strategy.
We may make use of the full off-resonance and multipolar resonant response of the high frequency system ({\it e.g.}, for a given capacitive tuning, resonance is achieved simultaneously at a single frequency between each inductive zero crossing). This yields the improved sensitivity illustrated by ``Multipole", which interestingly extends all the way to masses probed by ADMX.

\noindent
{\bf Discussion.---}In this Letter we demonstrate that the sensitivity of lumped-element axion detectors begins to break down relative to MQS sensitivity at frequencies over two orders of magnitude lower than the naive expectation.  
In the SM we show that solenoidal geometries give  similar behavior.

High quality factor lumped-element detectors remain promising for probing low-mass axions.
For example, an axion with $m_a = 0.5$ neV would have a decay constant at the supersymmetry GUT scale and be in range of DMRadio-GUT~\cite{Brouwer:2022bwo}, where $m_a r \lesssim 10^{-2}$ and thus the effects discussed in this work should not degrade the sensitivity relative to the expectation under the MQS approximation.
On the other hand, we show that additional challenges arise in maintaining the sensitivity of lumped-element detectors approaching their self-resonant frequencies.
Methods for mitigating these effects are actively being pursued~\cite{MethodsPaper}.
Nevertheless, for sufficiently low-loss detectors the accessible parameter space of lumped-element experiments may be extended well beyond their originally targeted frequency ranges. 

\section*{Acknowledgements}
This work grew out of insights gained as part of the ABRACADABRA Collaboration, and we thank Collaboration members 
J.~Ouellet and L. Winslow, in particular, for helpful discussions.
We also thank R. Henning for assistance with preliminary inquiries with \texttt{COMSOL}, J. Thaler and K. Zhou for collaboration on early stages of this work, A.~Niknejad for insightful discussions, and J. Thaler for carefully reviewing this manuscript. This work was supported by the Molecular Graphics and Computation Facility at UC Berkeley. J. Foster was supported by a Pappalardo fellowship. Y. Kahn is supported in part by DOE grant DE-SC0015655. J. N. Benabou and B. R. Safdi were supported in part by the DOE Early Career Grant DESC0019225. C.~P.~Salemi was supported in part by the National Science Foundation Graduate Research Fellowship under Grant No.~1122374 and by the Kavli Institute for Particle Astrophysics and Cosmology Porat Fellowship.

\bibliography{refs}

\clearpage

\onecolumngrid
\begin{center}
  \textbf{\large Supplementary Material for Detecting axion dark matter beyond the magnetoquasistatic approximation}\\[.2cm]
  \vspace{0.05in}
  {Joshua N. Benabou, Joshua W. Foster, Yonatan Kahn, Benjamin R. Safdi, and Chiara P. Salemi}
\end{center}

\twocolumngrid

%%%%%%%%%% Merge with supplemental materials %%%%%%%%%%
\setcounter{equation}{0}
\setcounter{figure}{0}
\setcounter{table}{0}
\setcounter{section}{0}
\setcounter{page}{1}
\makeatletter
\renewcommand{\theequation}{S\arabic{equation}}
\renewcommand{\thefigure}{S\arabic{figure}}
\renewcommand{\thetable}{S\arabic{table}}

\onecolumngrid

This Supplementary Material provides additional details and results for the analyses discussed in the main Letter.

\section{Impact of Readout Noise}
The large reactance realized in the vicinity of the self-resonance poles and the increased resistance associated with radiative losses have the effect of reducing both the signal and thermal noise currents in the pickup system, which scale like $1/Z(\omega)^2$, compared to the readout noise, which scales like $1/Z(\omega)$. This results in a greater dependence on the magnitude of readout noise in determining the axion sensitivity of lumped element detectors in the high frequency regime. To inspect the importance of readout noise, we determine the sensitivity of our fiducial toroidal detector as in Fig.~\ref{fig:sensitivity}, but now for $\eta_A = 1$, \textit{i.e.}, for amplified readout at the Standard Quantum Limit.

\begin{figure}[!ht]
    \centering
    \includegraphics[width=.6\columnwidth]{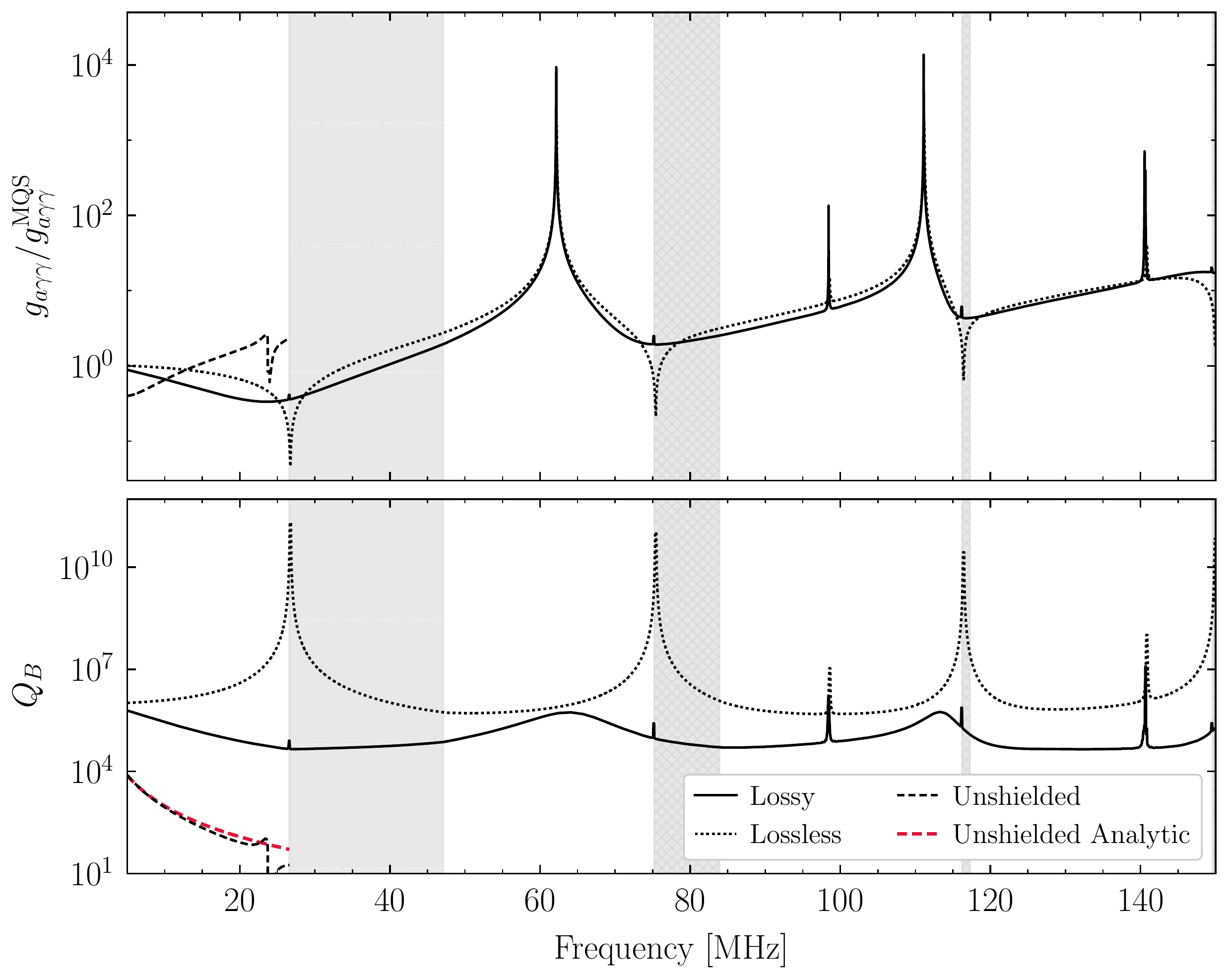}
    \caption{As in Fig.~\ref{fig:rel_sensitivity}, but for $\eta_A = 1$, \textit{i.e.}, with reduced readout noise.}
    \label{fig:rel_sensitivity_eta1}
\end{figure}

\begin{figure}[!h]
    \centering
    \includegraphics[width=.6\columnwidth]{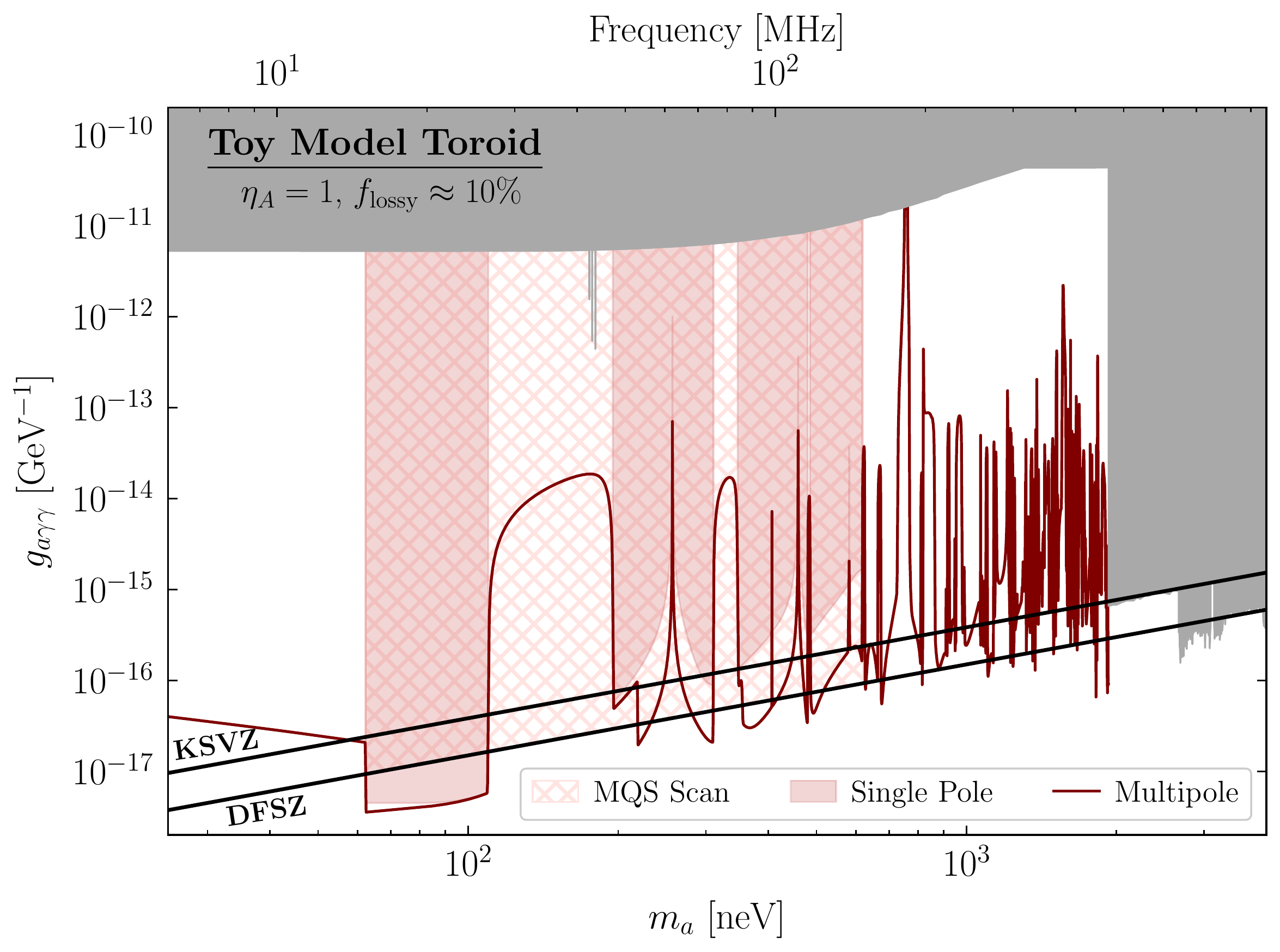}
    \caption{As in Fig.~\ref{fig:sensitivity}, but for $\eta_A = 1$, \textit{i.e.}, with reduced readout noise.}
    \label{fig:less_readout}
\end{figure}

In Fig.~\ref{fig:rel_sensitivity_eta1}, we show the relative sensitivity of a given resonant tuning to the axion-photon coupling for readout noise parametrized by $\eta_A = 1$, where the sensitivity is in fact enhanced at low frequencies prior to reaching the first self-resonant frequency, even for lossy detectors, before the large-scale trend of decreasing sensitivity with increasing frequency is restored. In Fig.~\ref{fig:less_readout}, we determine the sensitivity of an MQS scanning strategy for $\eta_A = 1$ with the high frequency response with and without incorporating multipolar and off-resonance sensitivity, as in the main Letter.

\section{Impact of the Lossy Material Fraction}
\begin{figure}[!h]
    \centering
    \includegraphics[width=.49\columnwidth]{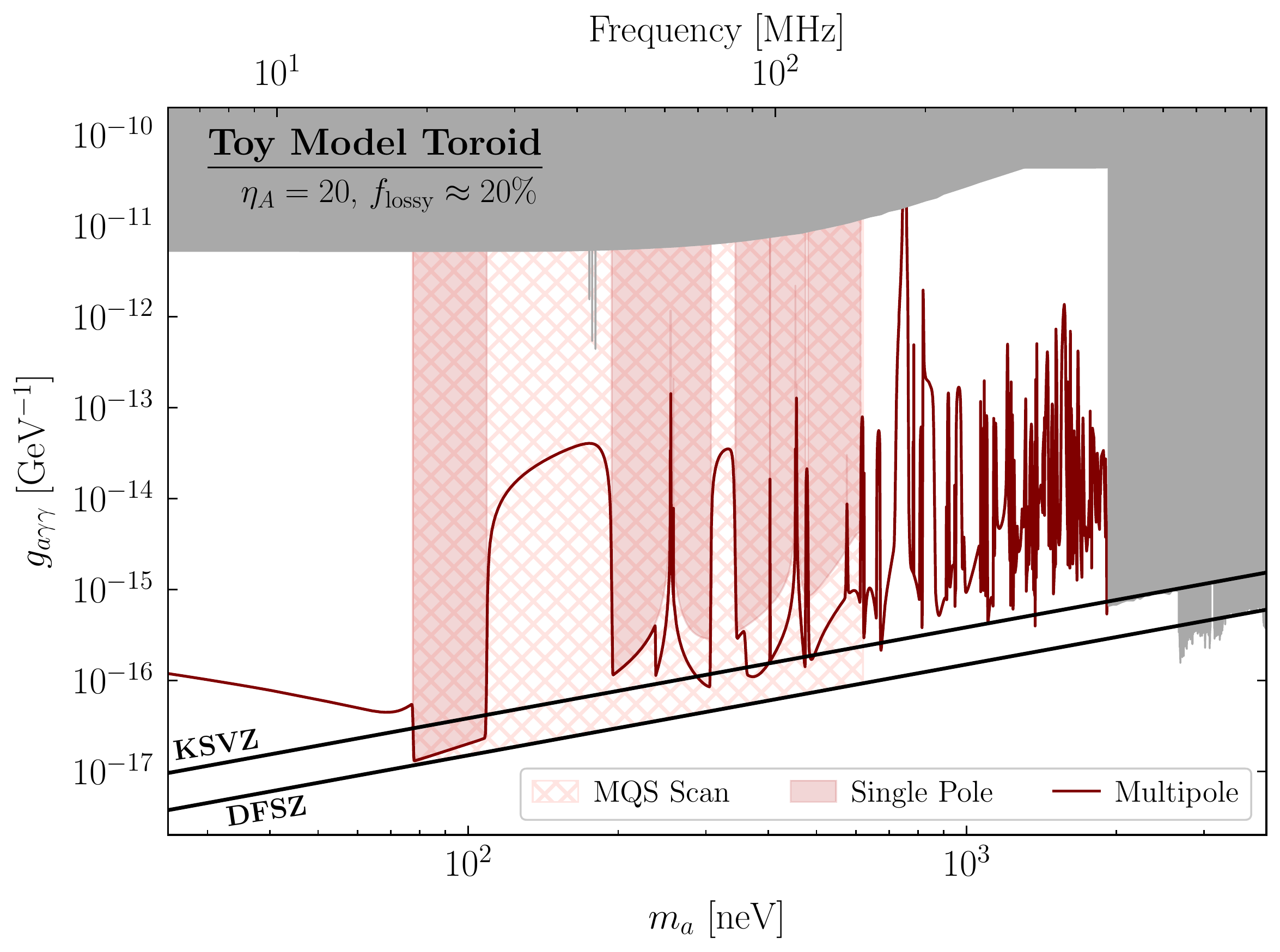}
    \includegraphics[width=.49\columnwidth]{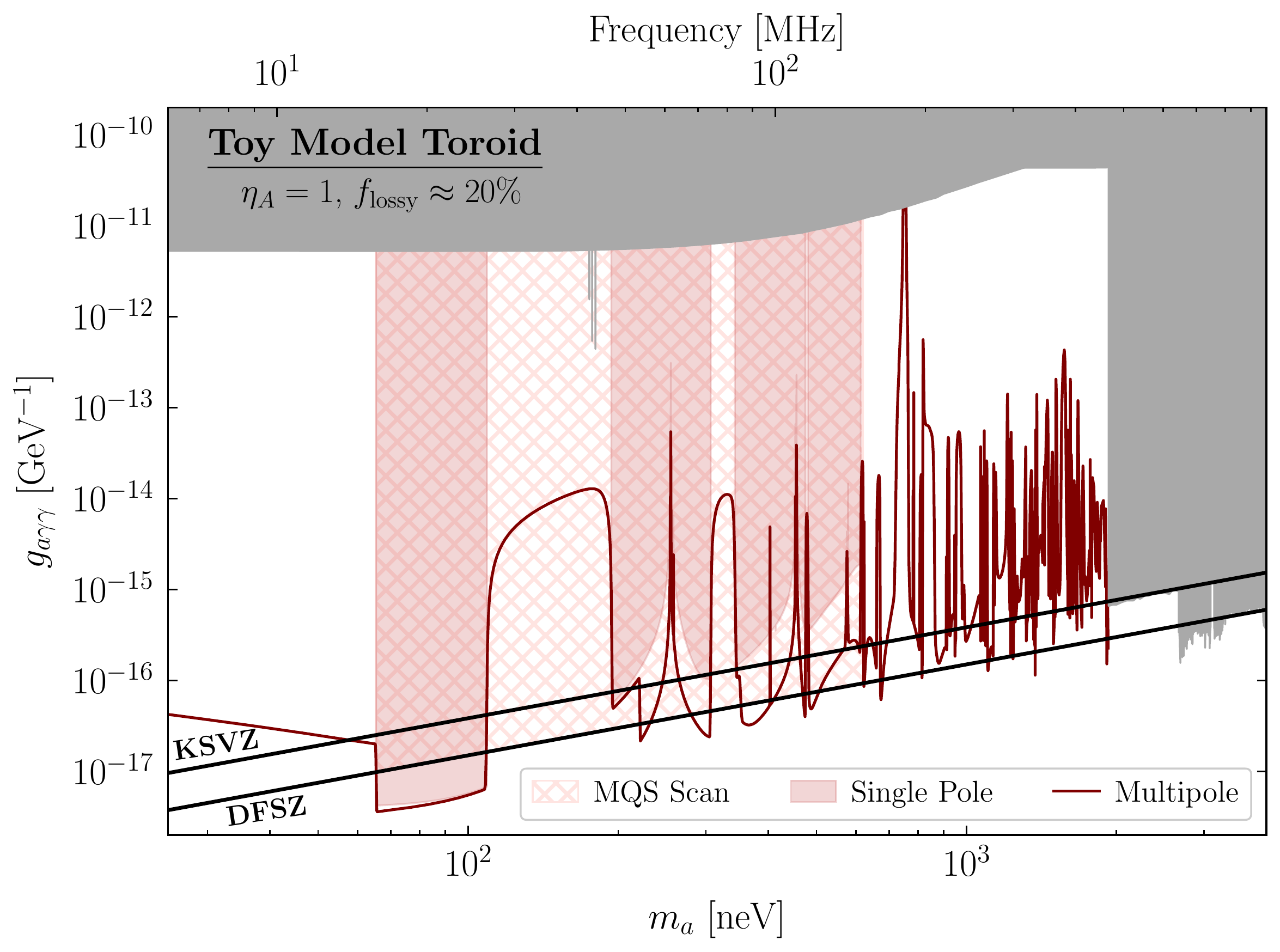}
    \caption{
    As in Fig.~\ref{fig:sensitivity} (left) and Fig.~\ref{fig:less_readout} (right), but with an increased lossy volume within the toroid of 20\%.}
    \label{fig:more_lossy}
\end{figure}

To test the impact of the fraction of lossy material within the toroid inner volume, we double the total lossy volume so that $f_\mathrm{lossy} \approx 20\%$.  As a reminder, this is the fraction of lossy material within the toroid volume and not within the shield volume; the latter fraction is much lower. The projected sensitivities associated with this increased lossy volume fraction for both $\eta_A = 20$ and $\eta_A = 1$ scenarios are shown in Fig.~\ref{fig:more_lossy}, with the limit-setting sensitivities slightly reduced.

\section{Sensitivity with a Solenoidal Geometry} 
\begin{figure}[!h]
    \centering
    \includegraphics[width=.49\columnwidth]{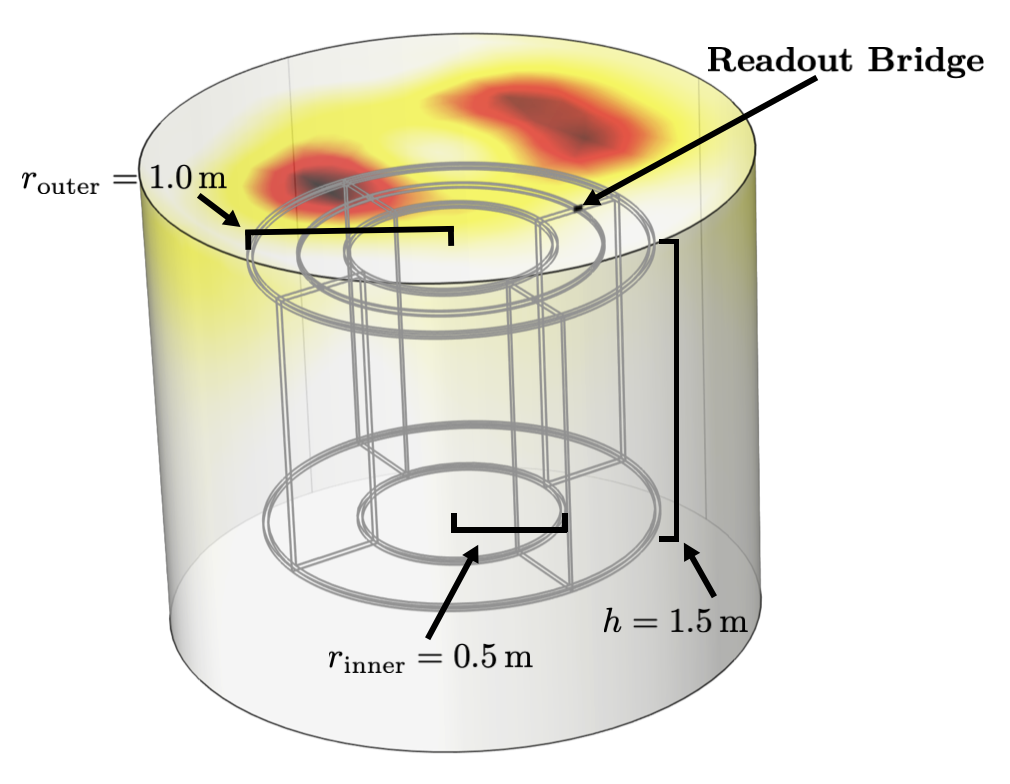}
    \includegraphics[width=.42\columnwidth]{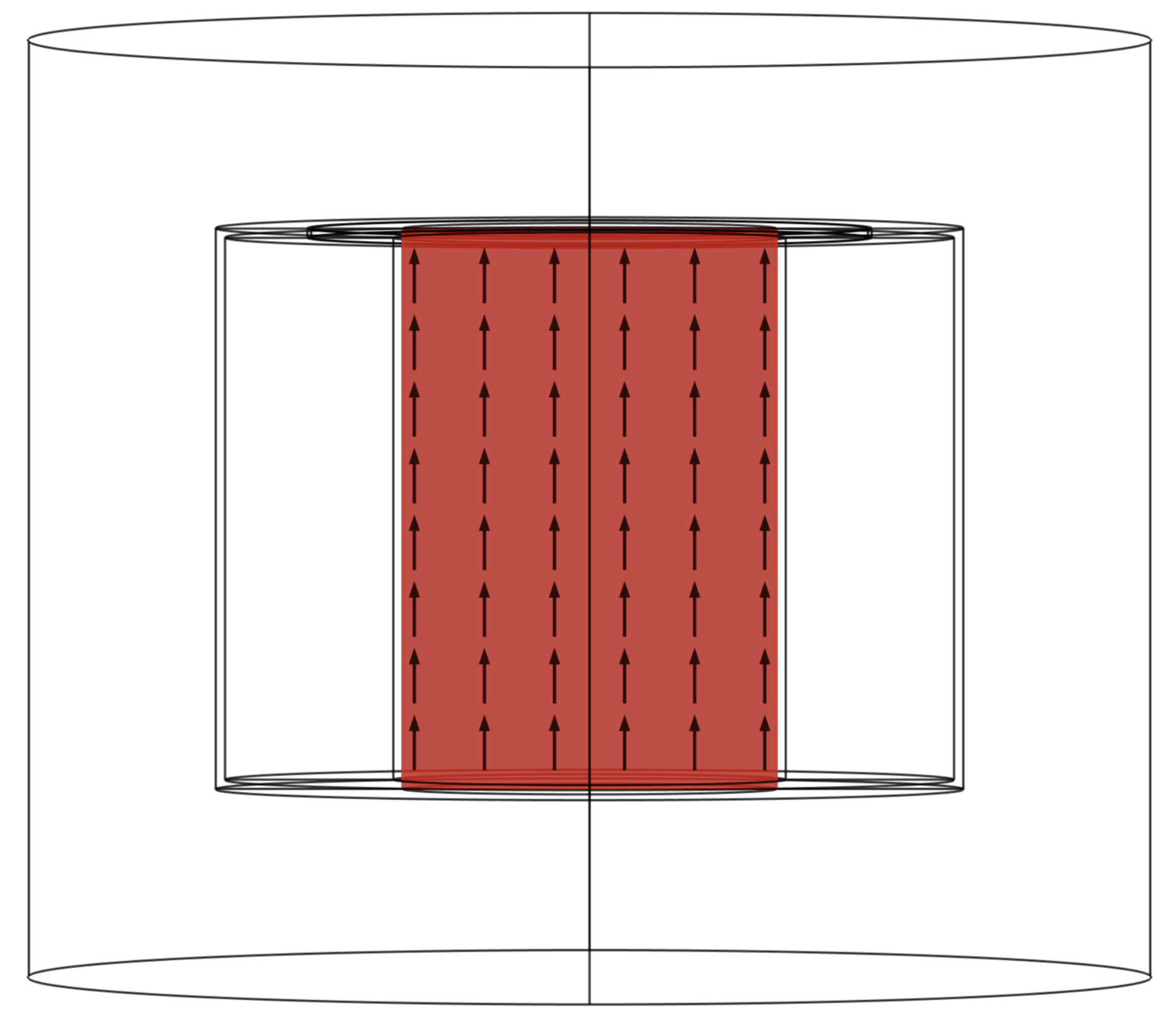}
    \caption{As in Fig.~\ref{fig:Toroid} (left), but for our solenoidal geometry. We additionally depict a side-view cross-section of the solenoid (right panel) with a heatmap and arrow plot that illustrates the simple static magnetic field assumed in this work. The magnetic field (red) is uniform and directed along the solenoid axis (arrows) within the enclosed volume and is zero elsewhere.}
    \label{fig:Solenoid}
\end{figure}

Thus far, we have focused exclusively on detection sensitivity employing a toroidal geometry. An alternate goemetry is that of a solenoid, which we study along the lines of our toroid study in this section. The simulated geometry is shown in detail in Fig.~\ref{fig:Solenoid} and is taken to be a cylindrical solenoid with $r_\mathrm{outer} = 2 r_\mathrm{inner} = 1\, \mathrm{m}$. The solenoid has height $h = 1.5\,\mathrm{m}$, with the circular top gapped by a $1\,\mathrm{cm}$ annulus across which we place our readout bridge and lumped port. The background 10 T magnetic field (and therefore axion effective current density) is taken to be uniform within the coaxial solenoid bore and directed along the solenoid axis. Outside, the field and axion effective current density are taken to be vanishing. In the lossy scenario, we line the exterior of the solenoid with TACHYON\textsuperscript{\textregistered} 100G
Ultra Low Loss Laminate and Prepreg with a total volume which is approximately $10\%$ the volume of the inner region.

\begin{figure}[!h]
    \centering
    \includegraphics[width=.7\columnwidth]{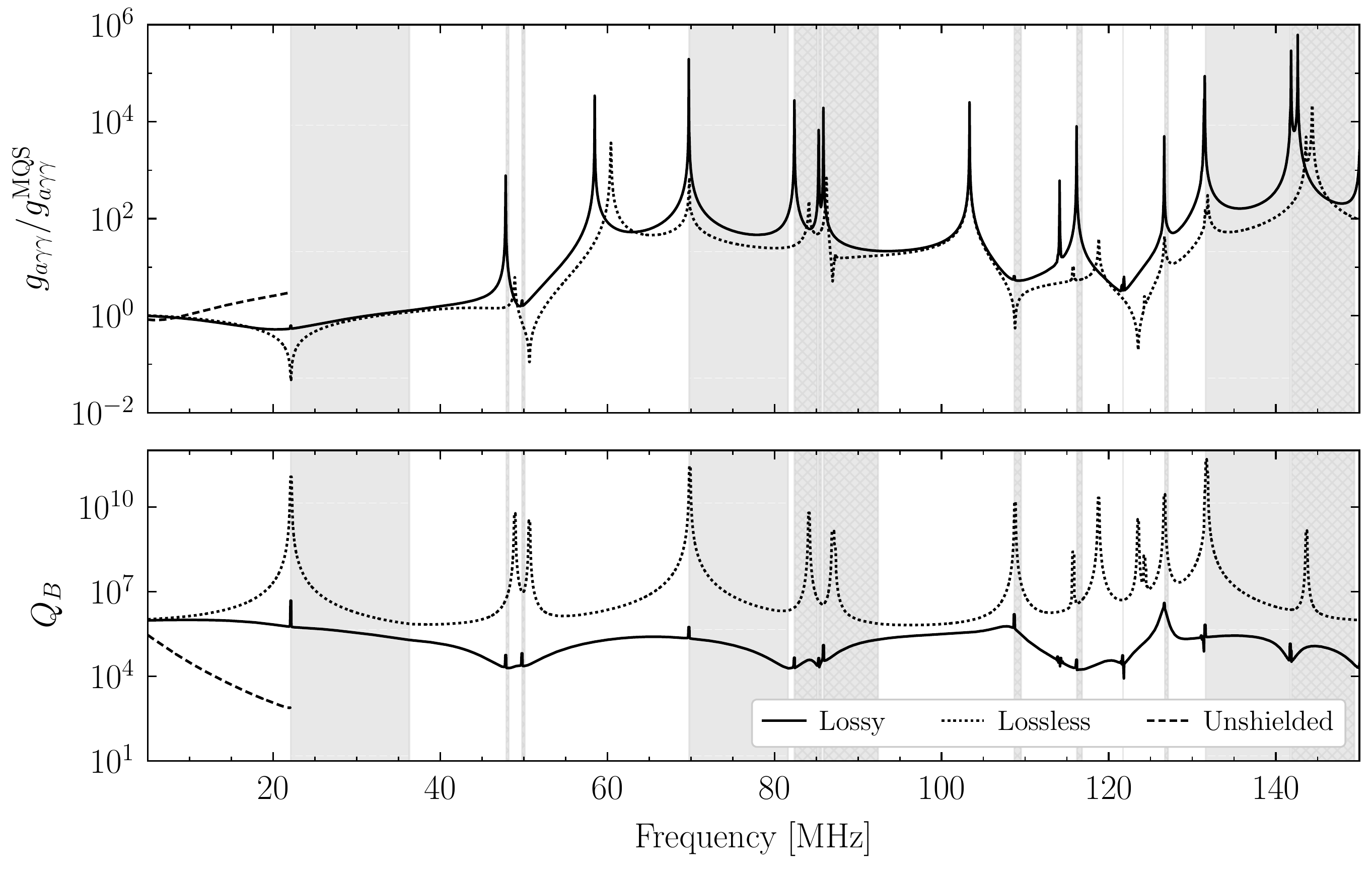}
    \caption{As in Fig.~\ref{fig:rel_sensitivity}, but for a solenoidal geometry.}
    \label{fig:rel_sensitivity_solenoid_eta20}
\end{figure}

In Fig.~\ref{fig:rel_sensitivity_solenoid_eta20}, we depict the relative sensitivity of the solenoidal detector in the high-frequency regime relative to the MQS expectations for the lossless, lossy, and unshielded scenarios, taking $\eta_A = 20$. (In the shielded scenarios the detector is in the middle of a 1.5 m radius and 2.5 m high cylindrical shield.) As in the case of the toroid, the signal and thermal current power are suppressed relative to MQS expectations, making readout noise an important factor in determining sensitivities at low frequencies that are beyond the MQS regime. However, even in the $\eta_A = 20$ scenario, the onset of high-frequency behavior prior to the first self-resonant frequency results in a slight enhancement of axion sensitivity in the lossy case.  This is in part because the solenoid has the advantage of being a less efficient radiator, given that the dipole moment associated with the axion-induced current density vanishes. 

\begin{figure}[!h]
    \centering
    \includegraphics[width=.49\columnwidth]{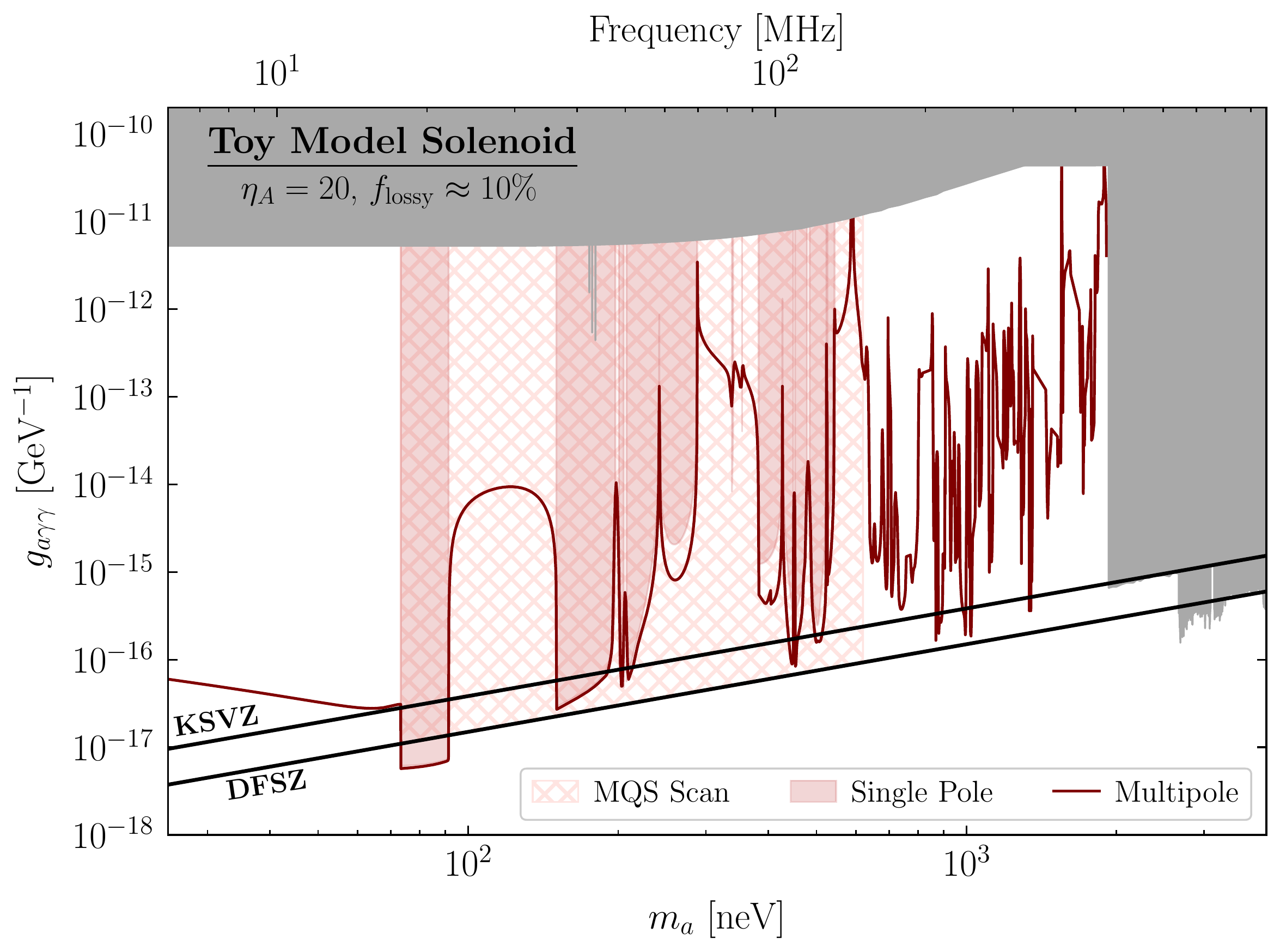}
    \includegraphics[width=.49\columnwidth]{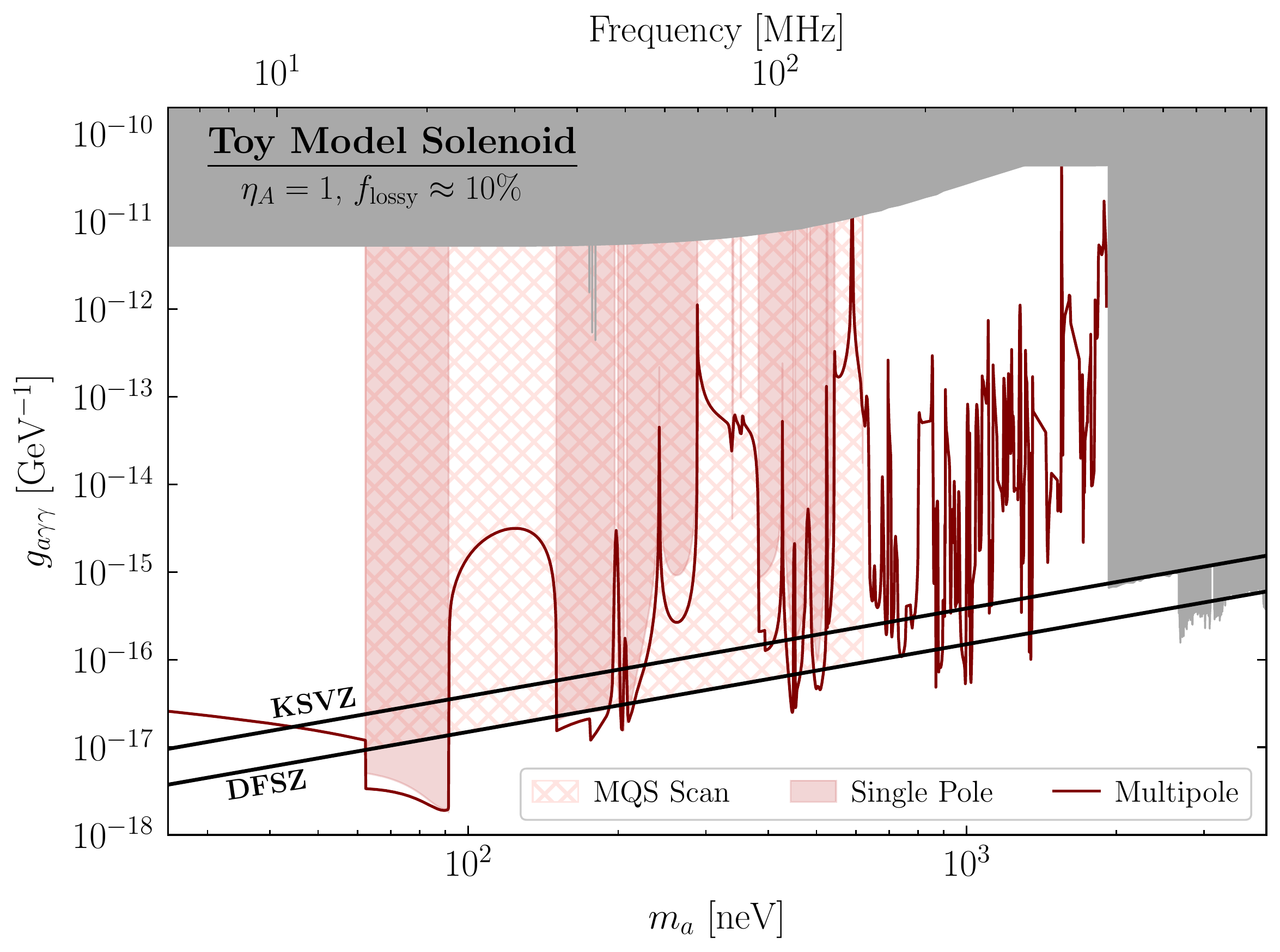}
    \caption{As in as in Fig.~\ref{fig:sensitivity} and Fig.~\ref{fig:less_readout}, but for our solenoidal geometry.}
    \label{fig:proj_senstivity_solenoid}
\end{figure}

In Fig.~\ref{fig:proj_senstivity_solenoid}, we depict the projected sensitivites of a MQS scanning strategy for the high frequency response with and without incorporating multipolar and off-resonance sensitivity for both $\eta_A = 20$ and $\eta_A = 1$. Here we adjust the scanning time to approximately $5.3$ years so that we achieve the same frequency coverage between our fiducial toroid and the solenoid in the $\eta_A = 20$ scenario under the MQS approximation.

\section{Broadband Readout Sensitivity}

As outlined in~\cite{Kahn:2016aff}, there are two general classes of readout systems for lumped-element detectors: broadband and resonant.  In this Letter we focus on resonant readout systems since they are generically more sensitive~\cite{Chaudhuri:2018rqn,Chaudhuri:2019ntz}, especially in the frequency range that will be probed by the upcoming DMRadio experiments.  However, it is interesting to also consider how the sensitivity of a broadband detector, which does not have additional lumped element components added to achieve resonance at a given frequency, is affected by beyond MQS approximation effects.  Broadband readouts are less affected by going beyond the MQS approximation because (i) there is no sense of a quality factor in the broadband case, which is otherwise degraded by radiative losses as we have described, and (ii) since there is no additional lumped element component added to the system, the fact that the reactance changes sign across $f_{\rm SRF}$ does not have practical implications. 

\begin{figure}[!t]
    \centering
    \includegraphics[width=.6\columnwidth]{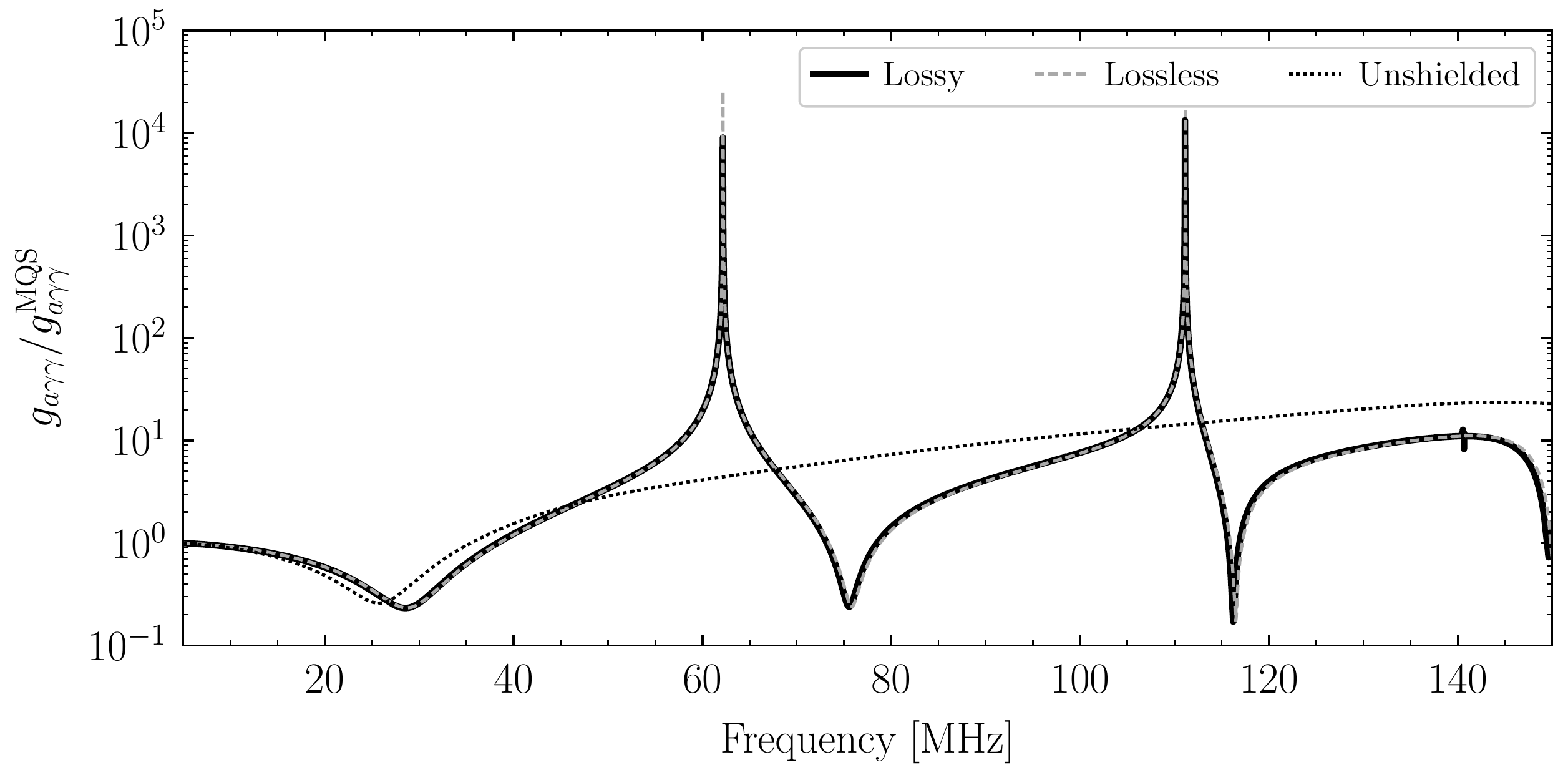}
    \caption{A comparison of the expected sensitivity of our fiducial toroid in broadband readout mode for our three boundary condition scenarios relative to the MQS expectations.}
    \label{fig:Toroid_Broadband_Rescale}
\end{figure}

\begin{figure}[!t]
    \centering
    \includegraphics[width=.49\columnwidth]{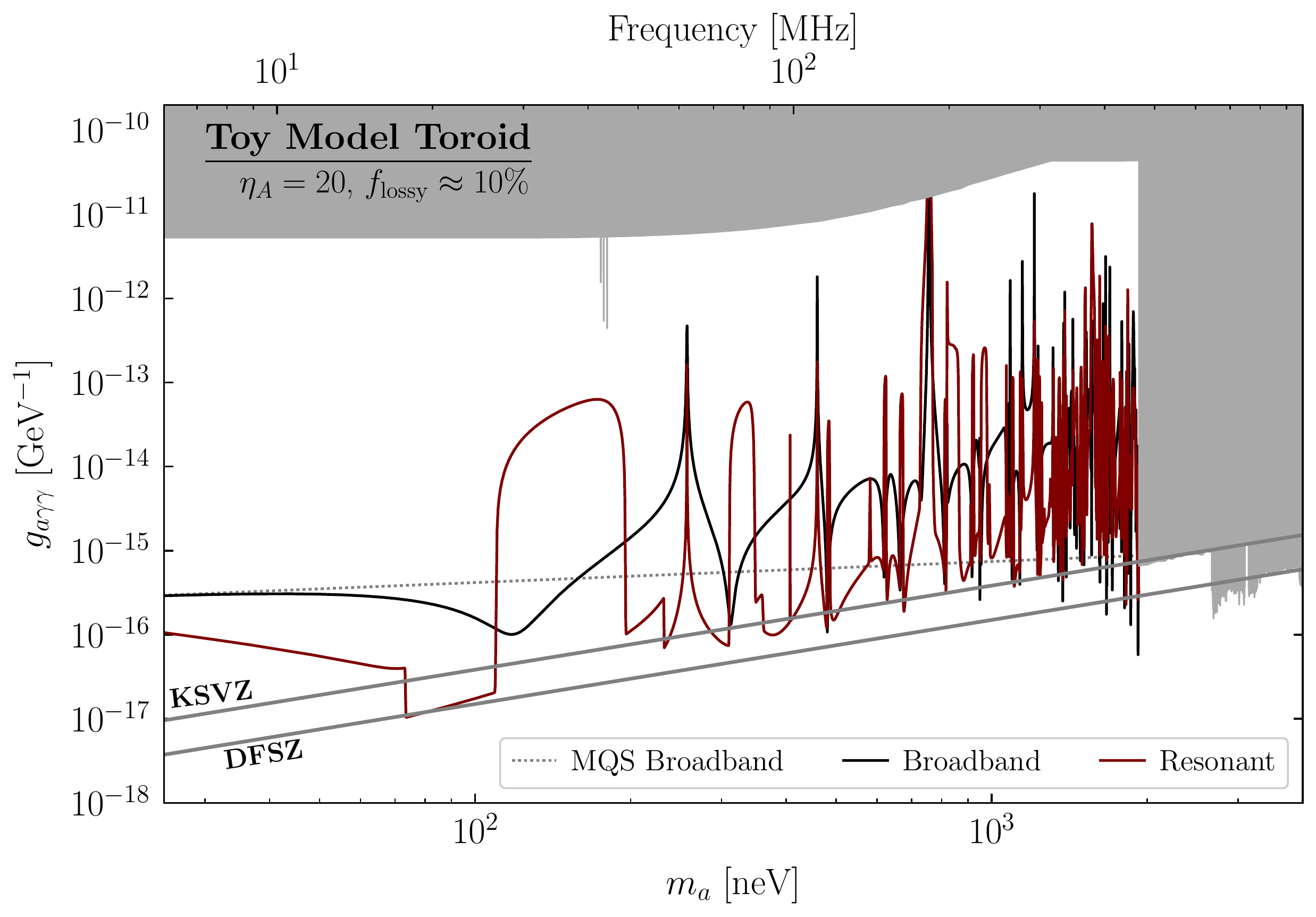}
    \includegraphics[width=.49\columnwidth]{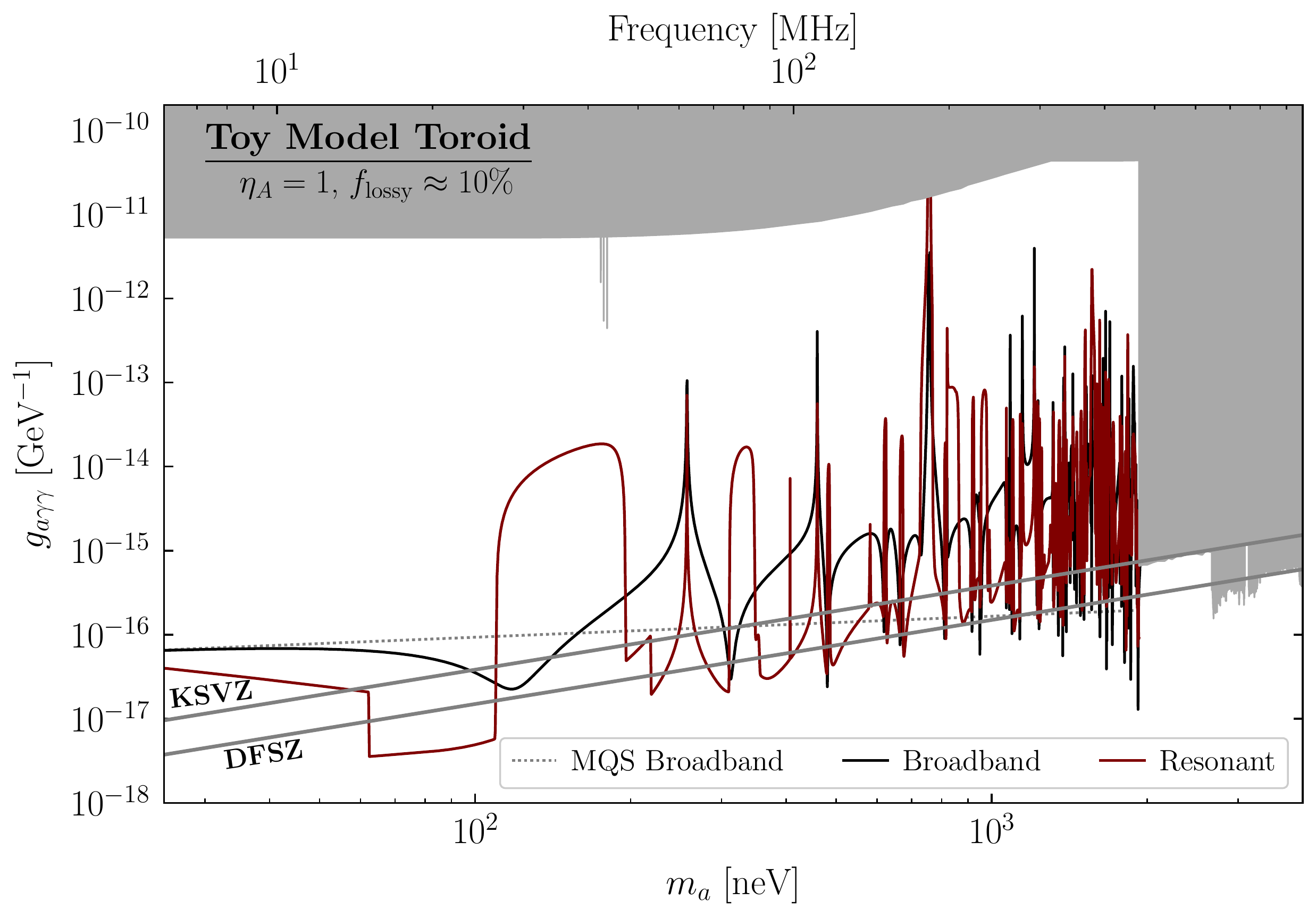}
    \caption{The projected sensitivity of our fiducial toroid to $g_{a\gamma \gamma}$ at 95\% confidence after one year of broadband scan time using the full high frequency response (black) and the MQS approximation (grey) for readout noise with $\eta_a = 20$ (left) and $\eta_a = 1$ (right). We compare to the sensitivity achieved with one year of resonant scanning with the same configuration accounting for the multipolar response in red.}
    \label{fig:Toroid_Broadband}
\end{figure}

\begin{figure}[!t]
    \centering
    \includegraphics[width=.49\columnwidth]{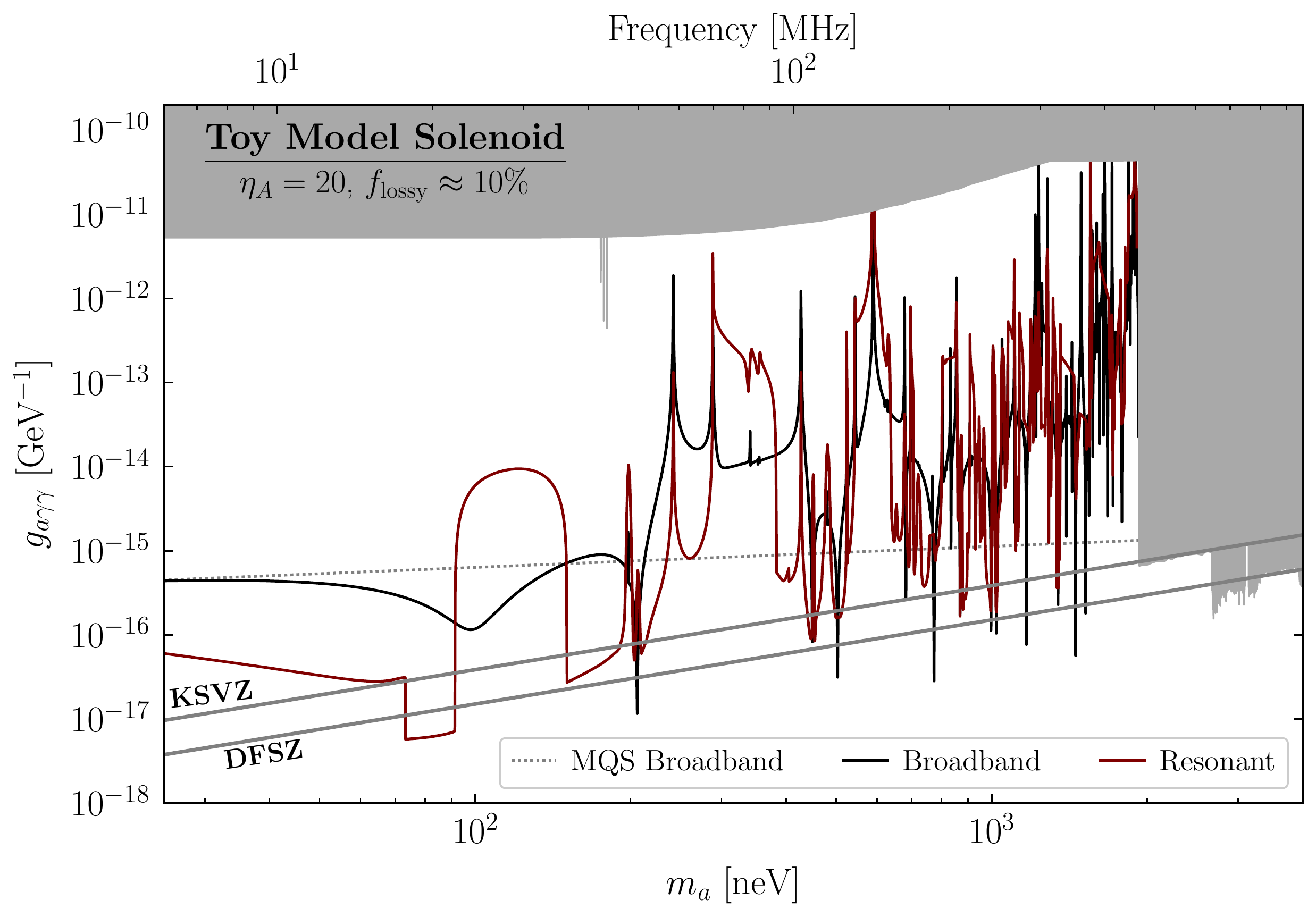}
    \includegraphics[width=.49\columnwidth]{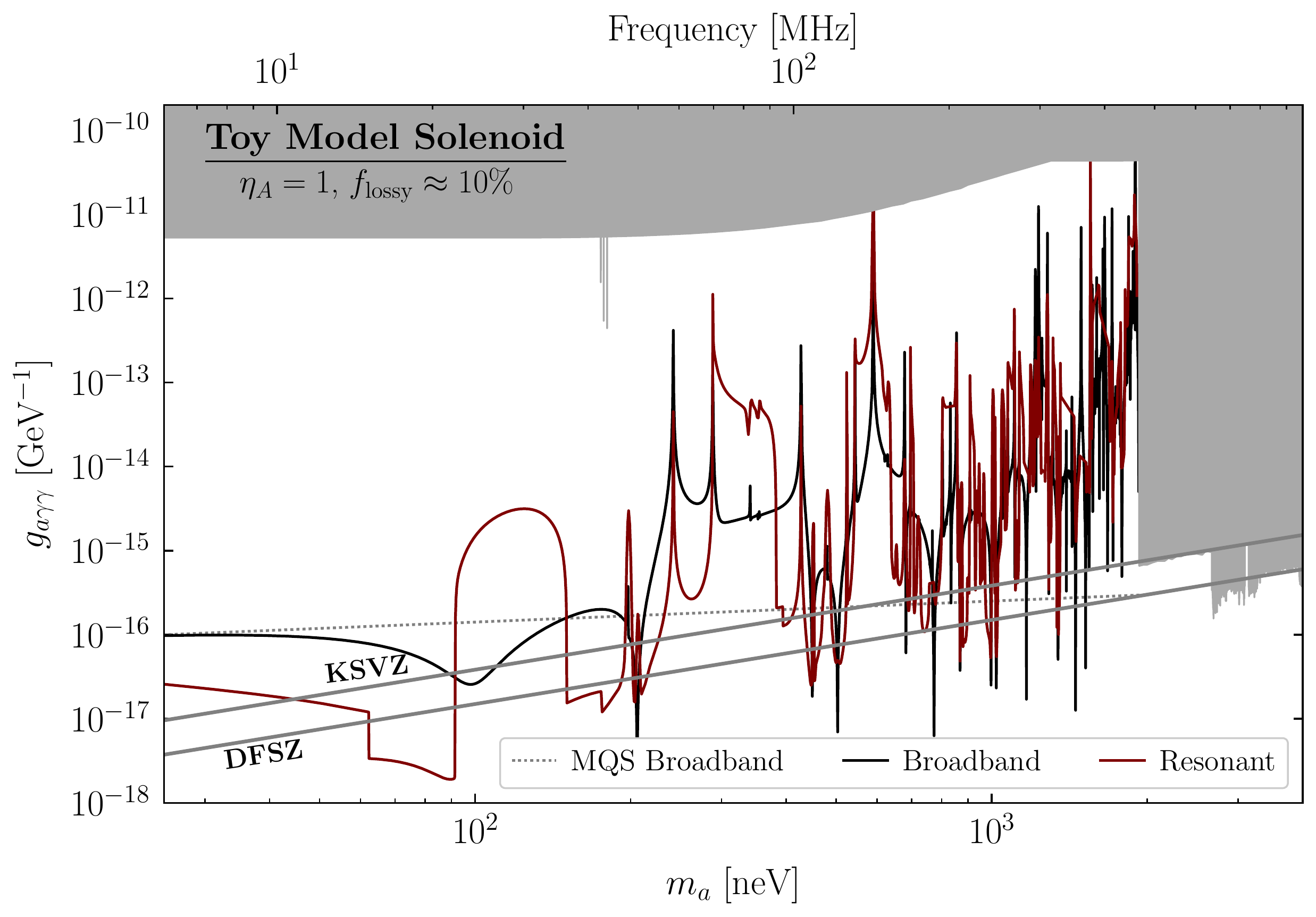}
    \caption{As in Fig.~\ref{fig:Toroid_Broadband}, but for the solenoidal geometry.}
    \label{fig:Solenoid_Broadband}
\end{figure}

In Fig.~\ref{fig:Toroid_Broadband_Rescale} we show how the sensitivity of our fiducial toroid is affected, with a broadband readout, when going beyond the MQS approximation relative to the expectation from the MQS approximation. We assume quantum-limited readout noise parametrized by $\eta_a$ and neglect contributions of thermal noise. Since we are operating a quantum limited readout, we choose the optimal readout configuration, in which the input coil that inductively couples the SQUID to the pickup contributes negligibly small impedance to the system. We operate the quantum-limited readout in the optimal configuration for broadband measurement in the MQS regime by taking $S_{VV}^\mathrm{BA} = \omega^2 L_{P}$ and $S_{II}^\mathrm{imp} = 1 / L_{P}$, where $L_\mathrm{P}$ is the low-frequency inductance of the pickup. (See~\cite{Chaudhuri:2018rqn} for details.) This figure should be directly compared to Fig.~\ref{fig:rel_sensitivity}.   As described above, relative to the resonant case, the sensitivity of the broadband readout is not severely affected by the MQS regime. In particular, the presence of lossy material does not affect the sensitivity so long as there is a shield.  Note that this implies that the results of the ABRA-10 cm experiment~\cite{Ouellet:2018beu,Salemi:2021gck}, which took data in broadband mode, are not affected by beyond-MQS effects.   Moreover, there is even a slightly increase in sensitivity near $f_{\rm SRF}$, relative to the MQS expectation, associated with greater noise suppression than signal suppression at the first self-resonance.  On the other hand, as we show in Figs.~\ref{fig:Toroid_Broadband} and \ref{fig:Solenoid_Broadband} for the toroid and solenoid, respectively, with different choices of $\eta_A$, the broadband readout still performs worse than the resonant readout, at most frequencies, even accounting for beyond-MQS approximation effects. 

\end{document}